\documentclass[rsi,floatfix,twocolumn,showpacs,superscriptaddress]{revtex4-1}
\usepackage{graphicx}
\usepackage{amssymb}
\usepackage{amsmath}
\usepackage{soul}
\usepackage[usenames]{color}

\ifx\pdftexversion\undefined
\usepackage[dvips]{hyperref}
\else
\usepackage{hyperref}
\fi
\hypersetup{
  colorlinks = true, linkcolor = blue
}

\begin{document}

\title{Characterization of a detector chain using a FPGA-based Time-to-Digital Converter to reconstruct the three-dimensional coordinates of single particles at high flux}
\author{F. Nogrette}
\affiliation{Laboratoire Charles Fabry, Institut d'Optique Graduate School, CNRS, Univ. Paris Sud, 2 Avenue Augustin Fresnel 91127 PALAISEAU cedex, France}
\author{D. Heurteau} 
\affiliation{F\'ed\'eration de Recherche LUMAT (DTPI), CNRS, Univ. Paris-Sud, Institut d'Optique Graduate School, Univ. Paris-Saclay, F-91405 Orsay (France)}
\author{R. Chang}
\affiliation{Laboratoire Charles Fabry, Institut d'Optique Graduate School, CNRS, Univ. Paris Sud, 2 Avenue Augustin Fresnel 91127 PALAISEAU cedex, France}
\author{Q. Bouton}
\affiliation{Laboratoire Charles Fabry, Institut d'Optique Graduate School, CNRS, Univ. Paris Sud, 2 Avenue Augustin Fresnel 91127 PALAISEAU cedex, France}
\author{C. I. Westbrook}
\affiliation{Laboratoire Charles Fabry, Institut d'Optique Graduate School, CNRS, Univ. Paris Sud, 2 Avenue Augustin Fresnel 91127 PALAISEAU cedex, France}
\author{R. Sellem} 
\affiliation{F\'ed\'eration de Recherche LUMAT (DTPI), CNRS, Univ. Paris-Sud, Institut d'Optique Graduate School, Univ. Paris-Saclay, F-91405 Orsay (France)}
\author{D. Cl\'ement} 
\affiliation{Laboratoire Charles Fabry, Institut d'Optique Graduate School, CNRS, Univ. Paris Sud, 2 Avenue Augustin Fresnel 91127 PALAISEAU cedex, France}

\begin{abstract}
We report on the development of a novel FPGA-based Time-to-Digital Converter and its implementation in a detection chain that records the coordinates of single particles along three dimensions. The detector is composed of Micro-Channel Plates mounted on top of a cross delay line and connected to fast electronics. We demonstrate continuous recording of the timing signals from the cross delay line at rates up to $4.1 \times10^{6}$ s$^{-1}$ and three-dimensional reconstruction of the coordinates up to $3.2 \times 10^{6}$ particles per second. From the imaging of a calibrated structure we measure the in-plane resolution of the detector to be 140(20) $\mu\rm{m}$ at a flux of $3 \times 10^{5}$ particles per second. In addition we analyze a method to estimate the resolution without placing any structure under vacuum, a significant practical improvement. While we use UV photons here, the results of this work apply to the detection of other kinds of particles. 
\end{abstract}

\maketitle

\section{Introduction} \label{lab_intro}
 
The detection of energetic particles, such as UV and X-ray photons or accelerated electrons and ions, is central to many research areas ranging from time-of-flight mass spectrometry \cite{Blavette1993} and diffraction experiments \cite{Xhu2011, Boll2014} to the monitoring of metastable atoms \cite{Vassen2010}. A widespread technique to image these particles is provided by the use of a Micro-Channel Plate detector (MCP) which allows time-resolved two-dimensional imaging. In the above applications both the resolution (in the two-dimensional image and along the time axis) and the maximum rate at which particles can be detected are of primary importance.  To obtain time-resolved two-dimensional images from the MCPs many methods have been developed  \cite{Jagutzki1998}. In particular cross delay lines offer the possibility to reach high fluxes while maintaining a good spatial resolution along the spatial axes \cite{Jagutzki02}. The capabilities of cross delay lines have been investigated in different contexts, in combination with MCPs \cite{Jagutzki02, Siegmund2005, Czasch2007, Berry2012, Keller2014}, photocathodes \cite{ Owens2010} and Gas Electron Multipliers (GEM) \cite{Guedes2003, Zhou2009}. 

When recording three-dimensional information about the impact of incoming particles from delay-lines, the electronics which digitize and store timing signals are of fundamental importance. In particular, not only should the MCP and the delay line allow high fluxes but the entire detection chain, ending with the storing of particle coordinates as digital data, must support these high rates. As the amount of data increases, both when lowering the time resolution to digitize the particle coordinates and when increasing the particle flux, the bottleneck in the detection chain is often related to the digital data. This limitation is either due to the time-to-digital converter (TDC) coding rate or due to the data transfer to a hard memory. In the works cited above, the maximum rates at which data is recorded is a few $10^5$ s$^{-1}$ \cite{Jagutzki02, Siegmund2005, Czasch2007, Zhou2009}. In \cite{Jagutzki02} the authors anticipate read-out rates up to $1\times10^6 \ \rm{s}^{-1}$  but without a demonstration. In addition, while there exist several commercial TDCs digitizing at rates higher than 1 MHz per channel, we are not aware of any experimental evidence of a complete detection chain -- ending with the reconstruction of 3D coordinates of incoming particles -- capable of acquiring data continuously at these rates.

In this work, we report on the characterization of a complete detection chain to record the three-dimensional coordinates of incoming particles. The coordinates of a single particle are obtained by combining the timing information from the four outputs of the cross delay line. We have developed a FPGA-based TDC that allows us to demonstrate  continuous acquisition at rates up to $4.1\times 10^6 \ \rm{s}^{-1}$ per channel and to reconstruct three-dimensional coordinates of particles at rates up to $3.2\times 10^6 \ \rm{s}^{-1}$, significantly higher than that previously reported \cite{Jagutzki02, Siegmund2005, Czasch2007, Zhou2009}. An in-plane spatial resolution of $140(20) \ \mu \rm{m}$ (averaged over the 80 mm diameter MCPs) has been measured at a flux of $3\times{10^5 \ \rm{particles}}$ per second by imaging a calibrated structure. This spatial resolution corresponds to a temporal resolution of 0.3 ns RMS on the measurement of time intervals. Building on the fact that the value of the resolution is not limited by the discretization of time inherent to the use of the TDC, we discuss a quantity that allows us to estimate the resolution in the absence of any calibrating structure. Using this quantity we investigate the performance of the detector at different incoming fluxes. We observe that the in-plane spatial resolution degrades with the rate of reconstructed particles \cite{Jagutzki02}. 

This paper is organized as follows. We first present in section \ref{Sec:TDC} the architecture and the performance of the novel FPGA-based TDC. The experimental set-up used to characterize the complete detection chain is described in section \ref{sec:ExpSetup}. In  section \ref{sec:ContinuousAcq}, we measure the count rates per output channel of the delay line and that of reconstructed particles with three-dimensional coordinates. In the last section \ref{sec:Resol}, we study the resolution of the detector using two different approaches. First, a direct measurement by imaging a calibrated structure is presented. Building upon this result, we introduce a quantity which yields, within the scope of our application, the spatial resolution and we investigate the influence of the flux of incoming particles on the measured resolution.

\section{Overview of the FPGA-based TDC}\label{Sec:TDC}

\subsection{Architecture of the FPGA-based TDC}
 
\begin{figure}[h]
\includegraphics[width=\columnwidth]{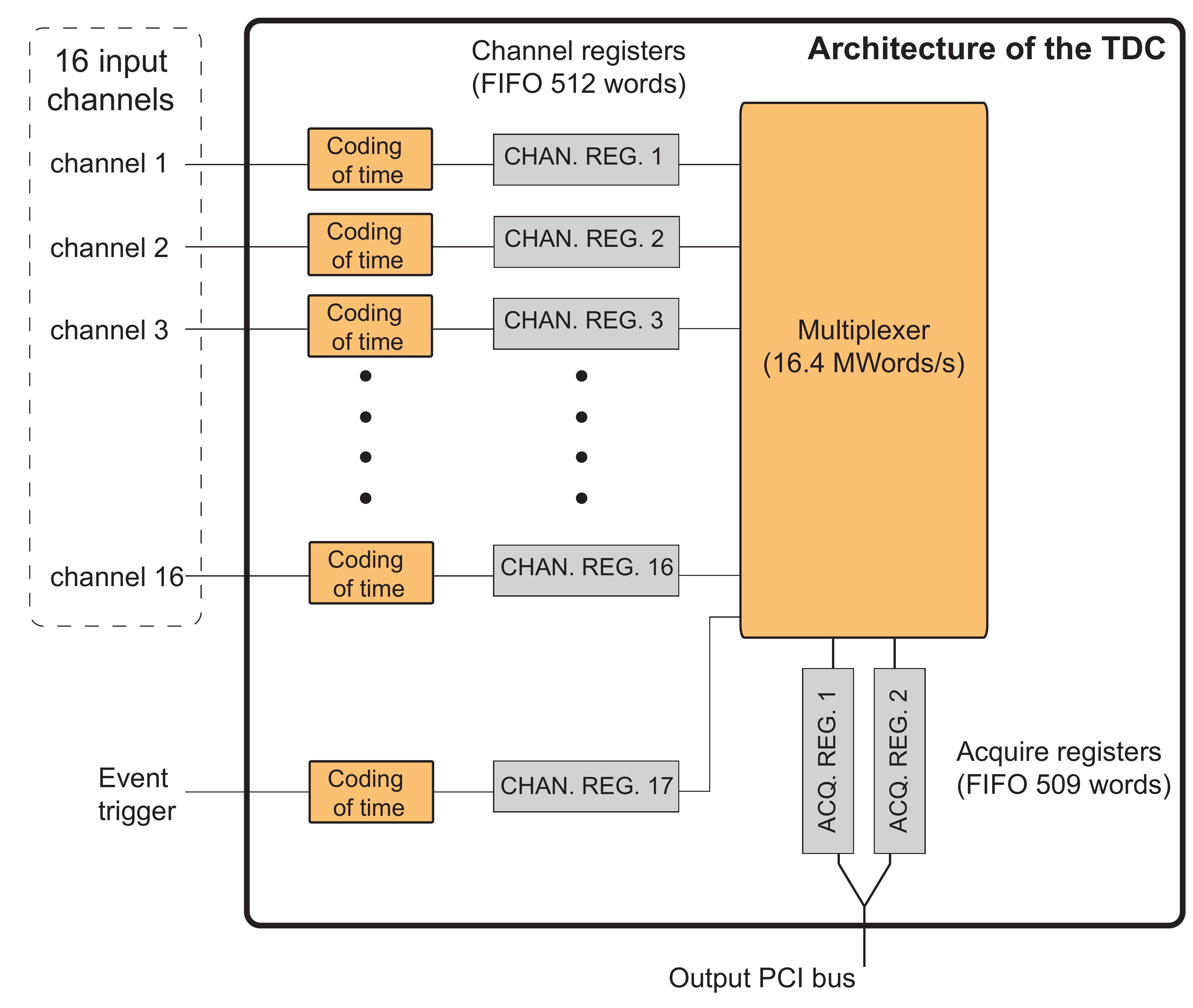}
\caption{Architecture of the FPGA-based TDC. There are up to 17 input channels, each of which stores the time associated to an incoming NIM pulse into a CHANnel REGister of depth 512 Words. The event trigger is time coded, stored and used as the reference from which time differences are evaluated off-line. The multiplexer is capable of treating 16.4 MWords/s and it stores the output data into two registers of depth 509 Words (ACQuire REGisters). Data is then stored into the computer RAM via the PCI bus. \label{Fig:FPGA}}
\end{figure}

The time to digital converter we have developed is a FPGA-based TDC \cite{ProductRef_of_TDC} using the Xilinx Virtex 4 technology. It features 17 independent input channels in which the arrival time of NIM signals can be digitized (see figure~\ref{Fig:FPGA}). The digitization of time relies on a 16-stage delay line locked to a 520 MHz reference clock, so that the delay of each elementary step does not depend on temperature, supply nor aging (with no need of operating calibration).  The duration of one elementary digitization step is equal to $t_{0}=120\ \rm{ps}$, corresponding to one Least Significant Bit (LSB). Moreover, thanks to the locked delay line, the reference frequency can be easily changed allowing one to modify the digitization step $t_{0}$ (up to $1.9\;$ns). This permits one, for instance, to increase the range of the TDC or to distinguish distortions due to the TDC from those due to the detector itself.

The time associated with the logical input signals (``stops") entering the TDC is digitized over 26 bits in a numerical word of 32 bits. The duration of the recording can be extended from 8 ms, corresponding to the 26-bit period, to an arbitrarily long value with a mechanism permitting off-line counting of an elapsed time period. The words are stored in a FIFO register (named CHANnel REGister (CHAN. REG.)) with a depth of $509$ words. Writing into the channel register results in an input dead-time of 2.5 ns.   

A coding sequence is enabled by an event trigger which is digitized in a similar word as the stops. The digitization of time is not deterministic because the TDC does not adjust the zero of its time base to the start of the acquisition. The TDC operates on a ``sliding scale" mode: the time base runs freely and the time associated with a given stop is the result of the difference between the two time codes corresponding to the stop and start signals. For an ideal TDC operating in sliding scale mode, the error in the determination of a duration is less or equal to 0.5 LSB, {\it i.e.} 60 ps RMS for the TDC we have developed. Conversely, the deviation of the mean value of many measurements of a given duration tends to zero with the number of measurements: in other words the TDC exhibits no ``Integral Non Linearity (INL)'' along its digitization range. Thus, the TDC does not induce any global distortion on the two-dimensional imaging in the MCP plane. 

The data words stored in the CHAN REG are multiplexed at 16.4 MWord/s  and stored alternatively in two acquisition FIFO registers with a depth of 509 words (ACQ. REG.). The multiplexing is done by periodically reading the channel registers so that a data word is retrieved from its channel register with the same probability independent of the channel. Finally, the acquisition registers are emptied at a 33 MWord/s rate and the data is transmitted through the PCI bus to the computer system. The stop times are digitally converted as long as the channel registers are not full: at high data rates, this implies that the number of converted stops from a given channel depends on the reading rate of the full data transfer. 

We have carefully optimized the depth of the ACQuire REGisters, on the one hand, and on the other hand, the PCI bus modes and wait states.  This makes it possible to reach the maximum throughput of the multiplexer. 

The duration of the acquisition can be fixed externally by a start pulse (event trigger) and an end pulse sent to the TDC. In the data acquisition program, a polling loop runs continuously until all the acquisition registers of a run are transmitted to the computer RAM memory. The storage on the hard-drive disk and the reconstruction algorithm are executed when the acquisition is complete. 
    
\subsection{Performance of the TDC}
 
To begin, we characterize the TDC by itself, independent of the rest of the detector. We first measure the Integral Non Linearity (INL ), the Differential Non Linearity (DNL) and the resolution of the TDC.  We check that the TDC exhibits no INL by sending pulses into a single input channel at a frequency of 20 MHz. The time interval coded by the TDC (50 ns) is stable to $\pm$2 ps, a lower limit given by the instrument we use. We also investigate the stability of the INL as a function of the temperature. The temperature of the TDC is measured and stabilized within $\pm 1$ degree Celsius ($10^6$ samples). In a range from 35 to 86 degrees Celsius, the INL is stable within the precision $\pm$2 ps of our measurement. 

\begin{figure}[h!] 
\includegraphics[width=\columnwidth]{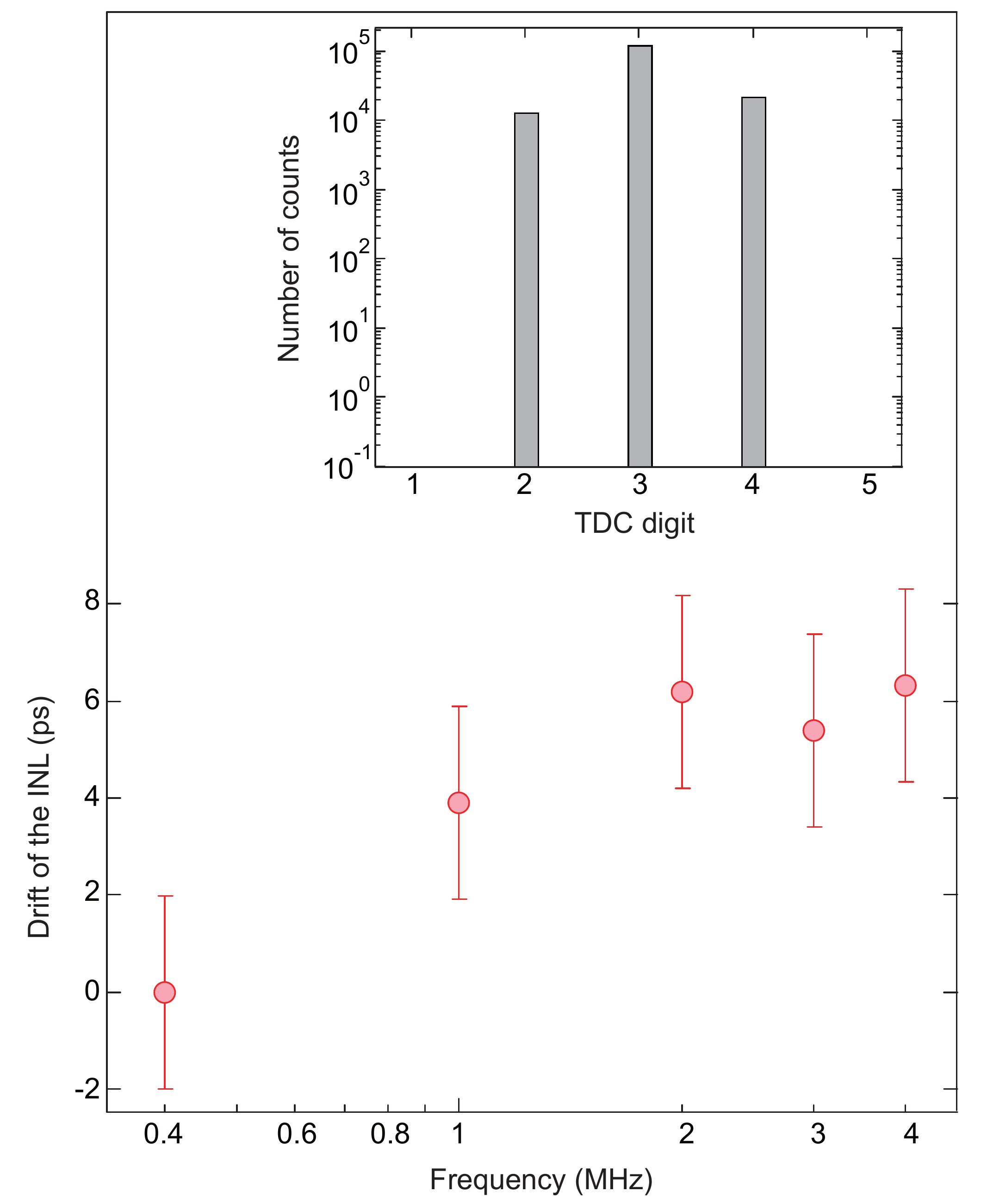}
\caption{Effect of the flux on the INL. The drift on the digitization of a fixed time interval (45 ns) between two separate channels of the TDC is plotted as a function of the repetition rate per channel. Inset: histogram of the counts on the TDC digit from a single measurement (flux of 4 MHz). }\label{Fig:TDC-flux}
\end{figure}

In addition, we send pulse pairs with a fixed time interval into two separate channels to investigate relative drifts, both as a function of the temperature and as a function of the acquisition rate. At room temperature and for incoming rates varying from 0.4 to 4 MHz (see Fig.~\ref{Fig:TDC-flux}) we measure a drift of $\sim$6 ps (on a time interval lasting 45 ns) negligible in comparison to the coding step $t_{0}=120~$ps.

\begin{figure}[h!] 
\includegraphics[width=\columnwidth]{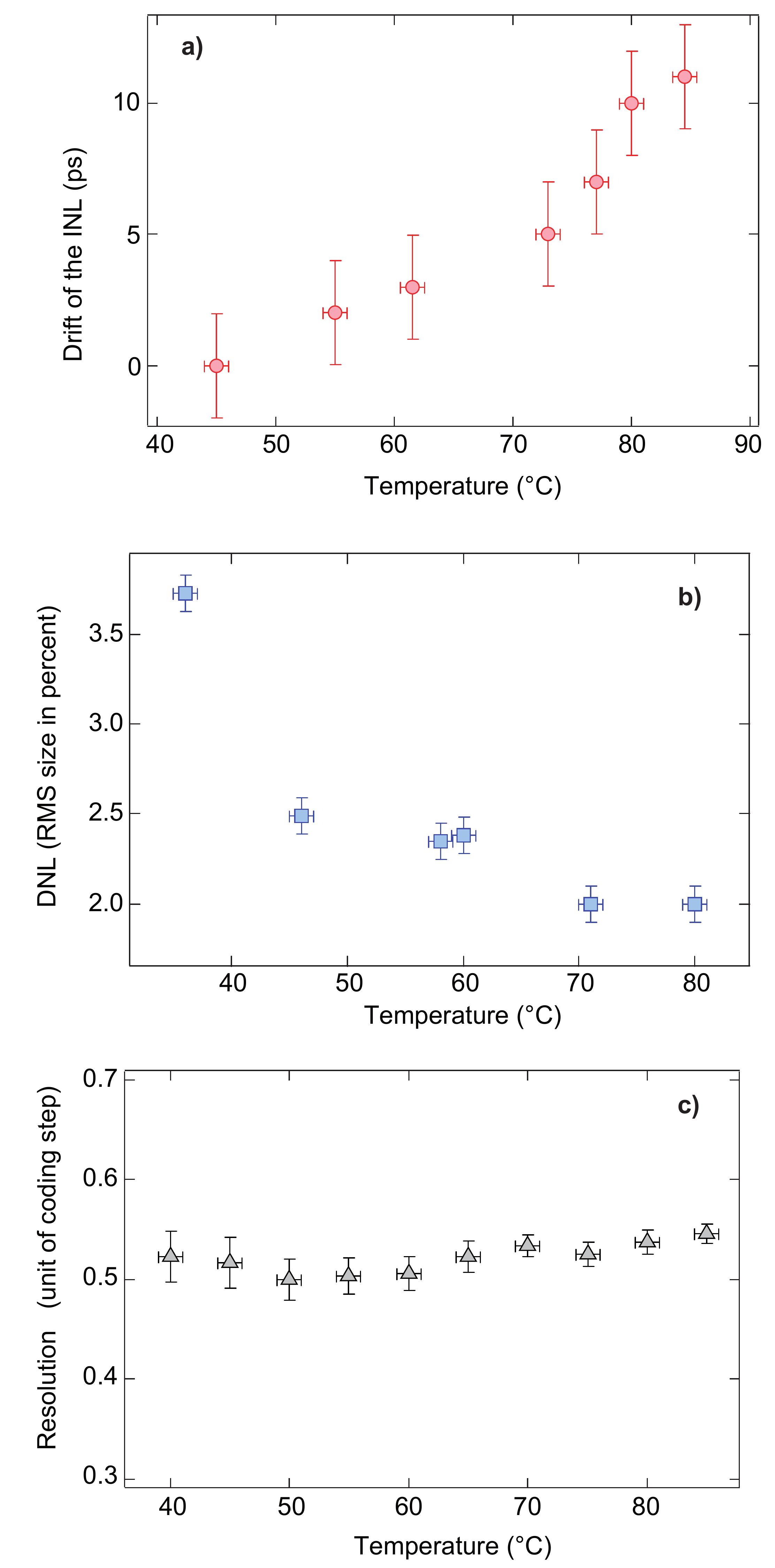}
\caption{Effect of temperature on the INL (a), DNL (b) and resolution (c) of the TDC. Temperature is measured and stabilized within $\pm 1$ degree Celsius (horizontal error bars). The vertical error bars reflect the level of accuracy of the measurement in the case of the INL (a) and the statistical deviation of the measurements in the case of the DNL (b) and resolution (c).}\label{Fig:TDC-temperature}
\end{figure}

In Fig.~\ref{Fig:TDC-temperature}~a) we plot the evolution of the measured time interval (over $10^6$ samples) and we observe a drift of $\sim 10$ ps over the range 45 to 85 degree Celsius, corresponding to a small fraction (8\%) of the coding step.

 The DNL quantifies the deviation of each digitization step from the ideal value; we measure a DNL (RMS value) less than 4\% of the coding step $t_{0}$ on each channel of the TDC ($4 \times 10^6$ samples). In addition we monitor the variation of the DNL ($10^8$ samples) on a given channel with temperature with the results plotted in Fig.\ref{Fig:TDC-temperature}~b. Over the temperature range we probe, the DNL is less than 4\%.

The resolution $\sigma_{0}$ of the TDC is defined as the standard deviation of the distribution of repeated measures of a given duration. In the case of a non-deterministic TDC with a zero DNL $\sigma_{0}$ is smaller than 0.5 LSB RMS \cite{Festa1985}. It is well known that a non-zero DNL slightly deteriorates the resolution \cite{Balla2014}. We measure a mean resolution $\sigma_{0}= 0.52(2)\ t_{0}$ over the 16 channels of the TDC, with a maximum value of $0.55\ t_{0}$ RMS in a specific channel. In addition, the resolution (measured over $10^6$ samples) is constant over a temperature range 40-85 Celsius degrees (see Fig.\ref{Fig:TDC-temperature}~c). 

To further characterize the TDC by itself, we measure the rate $\rho_{\text{stops}}$ at which stops are recorded with an 80 MHz digital pulse generator simulating the output pulses (NIM signals with 7 ns width) from the Constant Fraction Discriminator (CFD). The generator output is sent in parallel to 4 channels of the TDC to mimic the four outputs of the cross delay lines we use in the following, while the duration of the acquisition is determined by a second pulse generator.
We observe 2 regimes:
\begin{enumerate}
\item[(i)] when the amount of data does not exceed the depth of a CHAN. REG. (512 Words), we measure $\rho_{\text{stops}} = 80\ \rm{MWords/s}$ per channel. An acquisition at this rate is limited to a duration of $6.4$ $ \mu\rm{s}$.

\item[(ii)] when the amount of data exceeds the depth of a channel register, we measure a rate $\rho_{\text{stops}} = 4.12$ MWords/s per channel in agreement with the multiplexer speed (16.4 MWords/s) over 4 channels. 
\end{enumerate}

The role of the multiplexer is only visible in the latter regime.  Whatever the regime, the dead time of a channel input is low enough not to play any role here. In the second regime the transfer rate through the PCI bus (33 MWords/s) and the computing program polling speed are greater than the multiplexer speed. As a consequence the multiplexer sets the upper limit for the transfer rate of the TDC at $\rho_{\text{stops}} = 16.4$ MWords/s.

\section{Description of the test bench}\label{sec:ExpSetup} 

To characterize the detector as a whole we have set up a test bench consisting of an 80 mm diameter MCPs (Hamamatsu F1942-01 in chevron configuration) mounted above a cross delay line anode (Roentdek DLD80) and connected via vacuum feedthroughs to the electronics.

We illuminate the MCPs (in vacuum at a pressure of $2\times10^{-7}$ mbar) with UV photons from a deuterium lamp. The emission spectrum of the UV lamp (185 nm to 450 nm) is transmitted by a $\rm CaF_2$ viewport (cut-off wavelength $\sim $100 nm). The response of the MCPs strongly depends on the photon wavelength \cite{Eberhardt1979} with the consequence that the photons with the highest probability of detection have a wavelength $\lambda \approx 200$ nm corresponding to 6-7 eV. UV photons are illuminating the entire surface of the MCP and the flux of photons is adjusted with a variable diameter hole situated roughly 850 mm from the MCP. The MCPs are biased with 1kV voltage each and operate in the regime of gain saturation.
   
Outside vacuum, the electronics are composed of the following elements. First, the four signals at the end of the delay lines are decoupled from the high voltage and amplified. The double stage amplifiers have 1 GHz bandwidth with a typical total gain of 20. Second, the signals are transformed into logical NIM signals through four Constant Fraction Discriminators (CFD) (Amplifier and CFD circuits are Surface Concept ACU 5-6-8.1). The CFDs have a dead-time of 7 ns.  Finally the stop times are digitally converted with the FPGA-based TDC we have described above and are stored on the computer RAM.

\section{Continuous acquisition of data at high flux}
\label{sec:ContinuousAcq}

\begin{figure}[h] 
\includegraphics[width=\columnwidth]{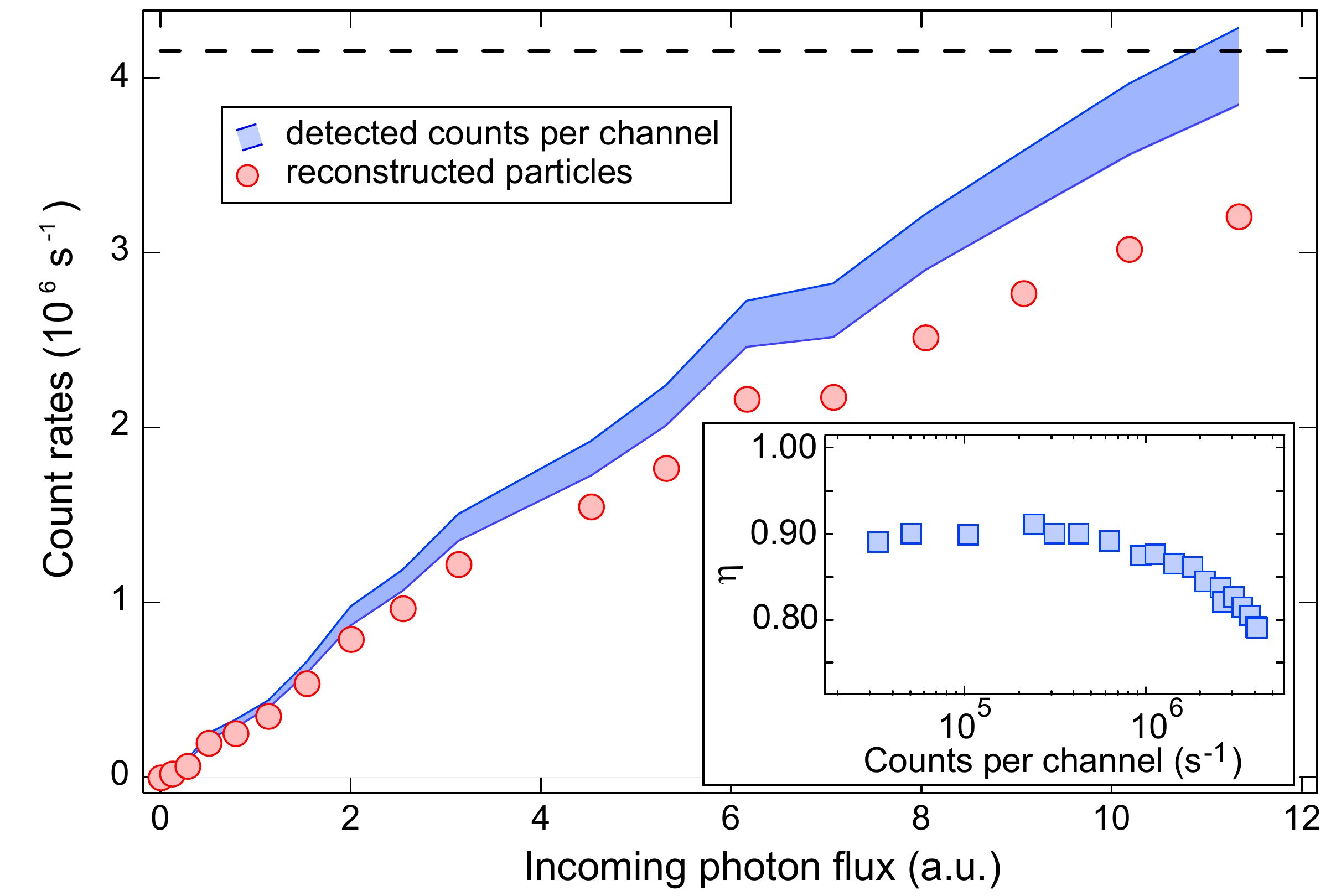}
\caption{Detected flux of counts (words) per output channel of the cross delay line (blue zone) and flux of reconstructed particles (red dots). The black dashed line corresponds to the maximum count rate per channel permitted by the TDC 16.4 MWord/s transfer rate. Inset: efficiency $\eta$ of the reconstruction process as a function of the incoming flux per channel (mean value of the four channels). }\label{Fig:CountRate}
\end{figure}

We are first interested in characterizing the count rates at which the detector ({\it i.e.} the complete chain of detection) works. We monitor two quantities, \textit{(i)} the count rate per channel and \textit{(ii)} the count rate of reconstructed particles. The count rate per channel is the number of stops per unit time recorded with the TDC on a given output channel of the cross delay line. The count rate of reconstructed particles is the number of particles per unit time with three-dimensional coordinates that one can correctly reconstruct based on the data acquired from the four delay-line channels. The event reconstruction is performed separately, and has no effect on the reconstructed count rate, therefore these computing resources are neither described nor characterized here.
The measurement of the count rates is shown in Fig.~\ref{Fig:CountRate} as a function of the incoming flux of UV photons. We have plotted the maximum and minimum rates measured on the four channels with a blue area in between the extremes. The number of stops per channel is the same to within $\pm6\%$ for each channel. The count rate per channel smoothly increases with the incoming flux of photons up to a maximal value of $\simeq 4.1\times 10^{6}$/s in average. This maximum value is limited by the multiplexing task of the TDC, the speed of which is 16.4 MWords/s. 
   
The particle reconstruction is based on the selection of electronic pulses from the four channels and thus depends on the entire chain of detection as well as on the computer algorithm. The rate of reconstructed particles we measure also smoothly increases with the photon flux up to a maximum value of $3.2 \times 10^{6}$ counts/s. As the photon flux increases, it deviates from the average count rate per channel. To quantify this behavior we plot in the inset of Fig.~\ref{Fig:CountRate} the ratio $\eta$ of the rate of reconstructed particles to the rate of counts per channel. As it can be seen in Fig.~\ref{Fig:CountRate}, the count rate per channel slightly deviates from a linear increase at high flux. This is expected from the piling-up of pulses in the electronics and from the local gain depression of the MCP stack at high incoming flux \cite{Edgar1992}. The reconstruction rate strongly depends on these losses per channel as a reconstructed particle requires the digitization of all four pulses exiting the delay-line anode. In addition, the reconstruction algorithm we implemented  considers all possible combinations of four digitized pulses and it does not induce any additional losses, a feature we checked.  We thus attribute the decrease in the reconstruction rate at high flux to the observed increasing losses per channel. Nevertheless the fraction of the reconstructed particles lies above 80\% over the entire dynamic range allowed by the TDC. We can compare these values with the performance in \cite{Siegmund2005} where 50\% losses were reported at a rate of $5\times10^5$ reconstructed particles.

Importantly, the results in Fig.~\ref{Fig:CountRate} were obtained while continuously acquiring data. With our long acquisition time (10 ms), a large number of transfers from the ACQuire REGisters is necessary (typically $>150$). This means that one can further extend the acquisition window at no cost in count rate, allowing us to qualify this acquisition as ``continuous". The duration of the acquisition is not limited by the architecture of the TDC itself but by the size of the RAM as the digital words are continuously stored in the computer. We have checked that measurements with longer acquisition times (100 ms, far above our final application needs), at the maximum rates yield identical results. In that later case, the number of transferred registers exceeds 3000. 

These high rates in continuous mode must be compared to the maximum achievable rate for the delay line with a 80 mm diameter MCP. It is well known that the detection efficiency of a MCP saturates when the output current becomes equal to a few percent of the strip current \cite{Wiza1979}. A rough estimate of the flux of incoming particles at which the MCP we use saturates (5-10\% of the strip current), yields typical values between $5-10\times10^6$ counts/s. Further improvement of the electronics may allow us to reach this limit.

\section{Resolution of the detector}\label{sec:Resol} 

In this section, we focus on determining the resolution of the entire chain of detection. We first present a direct measurement of the spatial in-plane resolution obtained by imaging a mask positioned slightly above the MCP. In the following section, we describe a measurement that allows one to evaluate the spatial resolution without the need for placing a mask above the detector. Throughout this section, we denote by $x$ and $y$ the spatial position of the incoming particle in the plane of the MCP surface. The position along the third axis is coded by the arrival time $t_{\text{a}}$ of the particle on the surface of the delay line. 

The recording of the timing pulses at the output of the cross delay lines, respectively $(t_{x,1},\, t_{x,2})$ for $x$ and $(t_{y,1},\, t_{y,2})$ for $y$, allows one to reconstruct the three-dimensional coordinates of an incoming particle.   The position $x$ along the $x$-axis is given by the time difference $(t_{x,1}-t_{x,2})/2$ with
\begin{eqnarray}
t_{x,1}=t_{a}+\frac{T^{\text{tot}}}{2}-\frac{x}{v_{x}} \nonumber \\
t_{x,2}=t_{a}+\frac{T^{\text{tot}}}{2}+\frac{x}{v_{x}}, \label{Eq:tx1-tx2}
\end{eqnarray}
where $T^{\text{tot}}$ is the propagation delay from one end of the delay line to the other end and $v_{x}=1.05$ mm/ns is an effective velocity describing propagation of the electronic pulses in the delay line \cite{RoentDek_Manual}. Here the origin of spatial coordinates is set at the center of the delay lines. The position $y$ along the $y$-axis is extracted similarly from the time difference $(t_{y,1}-t_{y,2})/2$. The coordinate $t_{a}$ along the third axis is determined from the time sums $\frac{1}{2}(t_{x1}+t_{x2})$ or $\frac{1}{2}(t_{y,1}+t_{y,2})$. From these considerations, it is clear that the precision with which time is digitized is central to the resolution of the detector. 
 
\subsection{Direct measurement of the in-plane spatial resolution}\label{subsec:DirectMeasResol} 

A straight forward method to determine the spatial resolution in the $x-y$ plane, {\it i.e.} the size of the point spread function in the plane of the detector, is to image an object with small structures onto the detector. We use a mesh mounted  $4\rm{\ mm}$ above the front face of the first MCP plate. The mesh is a $0.05\rm{\ mm}$ thick Cu-Alloy sheet with a honeycomb pattern. The center to center separation is $p= 0.95$ mm with a wall thickness $\delta\rm{p}=0.08\rm{\ mm}$.
 
\begin{figure}[h]
\includegraphics[width=\columnwidth]{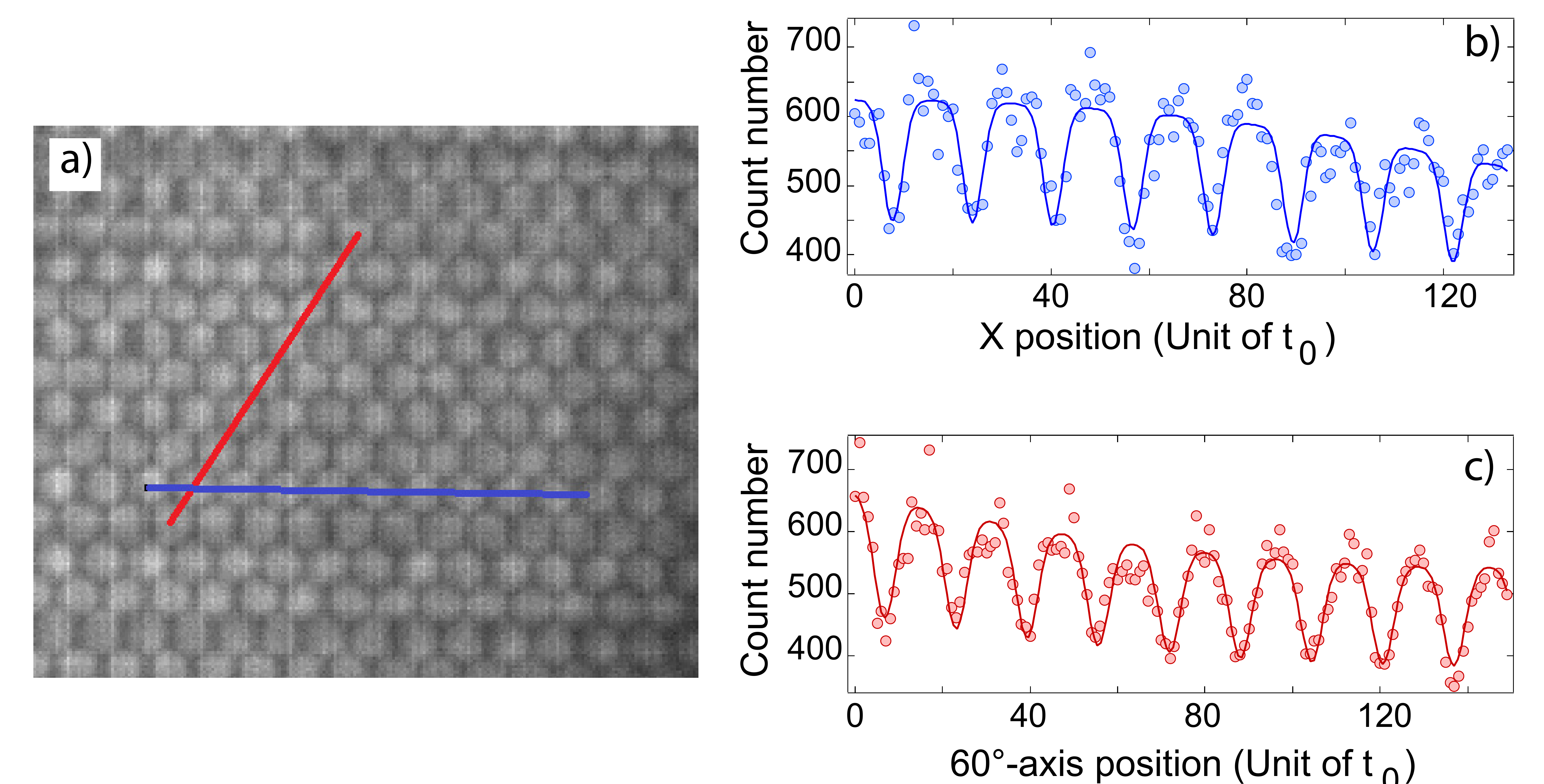}
\caption{ \textbf{a)} Image of the honeycomb mesh; the blue line is the zone of data for the profile along x and the red line for the 60 degree profile analysis. \textbf{b)} Spatial profile of the counts along the x axis, yielding a resolution $\sigma_{x}=2.0(2) \ t_{0}$. \textbf{c)} Spatial profile of the counts along the 60 degree axis, yielding a resolution $\sigma_{60}=2.5(2) \ t_{0}$.}\label{Fig:Grid}
\end{figure}

In figure \ref{Fig:Grid} a), we show an image taken in the presence of the mesh. In order to extract a spatial resolution from such an image, we plot the spatial profiles of the number of counts along two axes making an angle of 60 degrees (see Fig.~\ref{Fig:Grid} b) and c)). We assume that the shape of the dark regions -- corresponding to the walls -- is limited by the finite resolution of the detector. The profiles are fitted with a series of Gaussian functions, each with an identical RMS size and a separation given by the grid periodicity $p$ \cite{NoteSigmaX}. We define the in-plane resolution of the detector along a given axis by this RMS width. We obtain $\sigma_{x}=2.0(2) \ t_{0}$ along the $x$-axis and $\sigma_{60}=2.5(2) \ t_{0}$ along the axis making a 60 degree angle with the $x$-axis. The quantity $t_0$ is the digitization step. In spatial units, this corresponds to $120(20) \ \mu$m and $150(20) \ \mu$m respectively.  

\subsection{Estimating the resolution without imaging any structured object}\label{subsec:ExtractingRes} 

As discussed above, the time digitization of the sliding scale TDC sets a limit to the resolution with which a duration (evaluated as a time difference, such as $\Delta t=t_{x,1}-t_{x,2}$) can be measured. The spatial resolutions $\sigma_{x}$ and $\sigma_{60}$ we have measured lie well above this lower limit $\sigma_{0}\simeq 0.5 \ t_{0}$. Apart from the TDC itself, the sources contributing to the resolution are the electronic noise in the amplifiers and the  walk (jitter) of the discrimination process.  We assume all these sources of noise on a given channel can be considered independent from those of the other channels. We denote by $\sigma_{{\rm t, noise}}$ the RMS width of the time distribution associated with a single channel (and therefore the measurement of a single time, such as $t_{x,1}$). Since encoding a spatial position in the $x-y$ plane involves time differences as defined in Eq.~\ref{Eq:tx1-tx2}, we have
\begin{equation}
\sigma_{x}\simeq\frac{\sigma_{{\rm t, noise}}}{\sqrt{2}}, \label{Eq:SigX}
\end{equation} 
neglecting the contribution from the time digitization.

The spatial resolution can thus be extracted from a measurement of $\sigma_{{\rm t, noise}}$. We introduce a quantity $D$ to measure $\sigma_{{\rm t, noise}}$ over the entire surface of the detector defined as 

\begin{eqnarray}
D = (t_{x,1} + t_{x,2}) - (t_{y,1} + t_{y,2}).
\label{Eq:DefD}
\end{eqnarray}

\noindent $D$ is derived from the four times measured from the delay lines.  The RMS size $\sigma_{D}$ of its distribution depends on the noise contributions from the four different detection channels $\sigma_{{\rm t, noise}}$. 
Here, up to a constant geometrical factor, $t_{x,1}+t_{x,2}-2 t_{a}$ and $t_{y,1}+t_{y,2}-2 t_{a}$ are equal to the propagation time from one end of the delay line to the other end (see Eq. \ref{Eq:tx1-tx2}). For perfectly identical delay lines, cables and electronics, these two quantities are equal and $D(x,y)=0$. In practice the situation is more complicated and the spatial distribution $D(x,y)$ exhibits an offset and a gradient \cite{Hoendervanger2013}. In order to extract a position-independent RMS size of the $D$ distribution we proceed as follows. At a given position in the image plane $(x,y)$, we record local histograms of the $D(x,y)$ values from which we extract a local mean value $\langle D \rangle (x,y)$. The quantity $\sigma_{D}$ is defined as the RMS size of the distributions of $D(x,y)-\langle D \rangle (x,y)$ summed over the entire surface of the detector.

\begin{figure}[t!]
\includegraphics[width=\columnwidth]{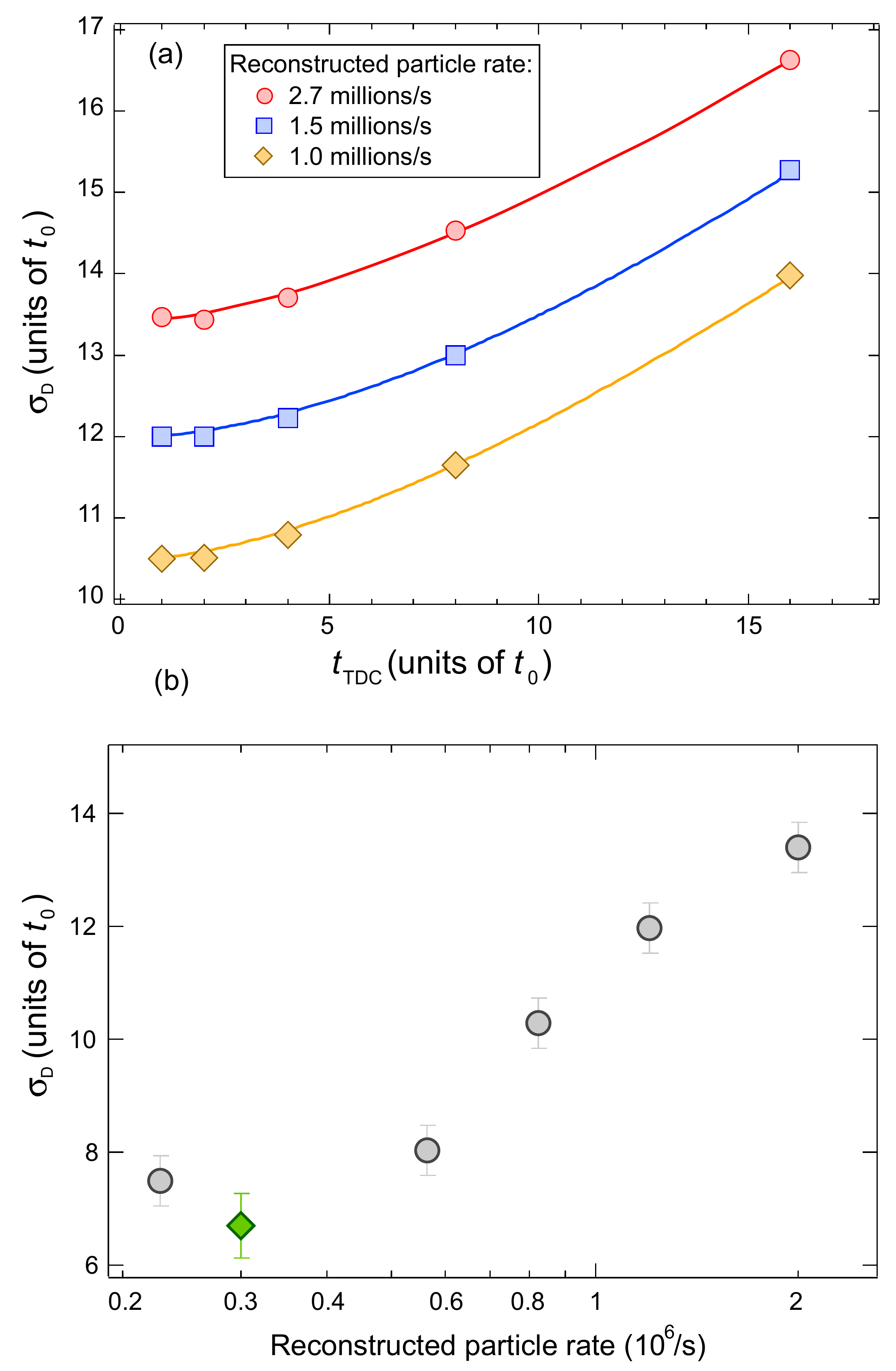}
\caption{{\bf a)} Plot of $\sigma_{D}$ as a function of the TDC digitization step $t_{\text{TDC}}$ for three different fluxes of incoming particles. All quantities are normalized to the elementary step $ t_{0}=120$ ps. {\bf b)} Variation of $\sigma_{D}$ (circles) using the minimum digitizing step ($t_{\text{TDC}}=t_{0}$) versus the incoming flux (expressed in reconstructed atom rate). The error bars reflect the variations of $\sigma_{D}$ as the pixel size is changed. The green diamond is the value of $\sigma_{D}$ obtained from the measured spatial resolution $\sigma_{60}$ (Eq.\ref{Eq:ResFromSigmaD}). }\label{Fig:SigD}
\end{figure}

The direct measurement of the spatial resolution presented in the previous section suggests that the contribution of electronic noise is larger than that from the time digitization. Here we are interested in investigating these relative contributions to the quantity $\sigma_{D}$. To do so, we measure $\sigma_{D}$ as a function of the elementary digitization step $t_{\text{TDC}}$ of the TDC that we artificially change by removing one LSB at a time in the 32-bit word coding time. We have increased $t_{\text{TDC}}$ from its lower value $t_{\text{TDC}}=t_{0}$ up to $t_{\text{TDC}}=16 \ t_{0}$. In Fig. \ref{Fig:SigD} a) we plot the measured value $\sigma_{D}$ as a function of $t_{{\text TDC}}$ and for different fluxes of incoming particles. The pixel size, over which $\langle D \rangle (x,y)$ is evaluated, is identical for all the $t_{\text{TDC}}$ values ($16 \ t_{0}$). 

We observe that $\sigma_{D}$ increases as a function of $t_{\text{TDC}}$ as expected since the digitization error increases with the coding step $t_{\text{TDC}}$ of the TDC. In addition, a large offset value in $\sigma_{D}$ is clearly visible at low $t_{\text{TDC}}$ values, thus confirming that the contribution of the uncorrelated noise $\sigma_{{\rm t, noise}}$ dominates that from the digitization error. Finally, we note that the quantity $\sigma_{D}$ depends upon the incident flux: the larger the flux the larger $\sigma_{D}$ (see below).

The contributions to the RMS width $\sigma_{D}$ from {\it (i)} the time digitization of the TDC and {\it (ii)} the uncorrelated sources of noise that we have defined previously, add independently, leading us to write
\begin{equation}
\sigma_{D}^2= \sigma_{q}^2(t_{{\rm TDC}}) +4 \ \sigma_{{\rm t, noise}}^2 \label{Eq:sigD}
\end{equation}
where the first term corresponds to the error coming from the time digitization and the second term results from the independent sum of the contributions of the electronic noise from the four input channels. Writing an explicit expression for $\sigma_{q}(t_{{\rm TDC}})$ as a function of $t_{0}$ is not straightforward without further assumptions, but an explicit expression is not required here. 

At low $t_{\text{TDC}}$ values where the saturation of $\sigma_{D}$ is clearly visible (see Fig.~\ref{Fig:SigD}a), the values of $\sigma_{{\rm t, noise}}$ largely exceed that given by the digitization error, whatever the incoming flux of particles. This indicates that the resolution is limited by independent noise in the different channels of the detection chain. For the resolution $t_{\text{TDC}}=t_{0}$, we conclude from our measurements (see Fig.~\ref{Fig:SigD}) and Eq.~\ref{Eq:sigD} that $\sigma_{D} \simeq 2 \ \sigma_{{\rm t, noise}}$. 

These considerations allow us to relate the quantity $\sigma_{D}$ to the spatial resolution (see Eq.~\ref{Eq:SigX}). We introduce a spatial resolution $\sigma_{\text{res}}$ for the detector defined by
\begin{equation}
\sigma_{\text{res}}=\frac{\sigma_{D}}{2\sqrt{2}}.\label{Eq:ResFromSigmaD}
\end{equation}

\noindent The resolution $\sigma_{\text{res}}$ is a combination of the spatial resolution along the two axis, $\sigma_{x}$ and $\sigma_{y}$ since, using Eq.~\ref{Eq:DefD}, one can write $\sigma_{D}^2=4(\sigma_{x}^2+\sigma_{y}^2$) and thus $\sigma_{\text{res}}^2=(\sigma_{x}^2+\sigma_{y}^2)/2$. When the resolutions along the $x$- and $y$-axis are equal, $\sigma_{x}\simeq \sigma_{y}$, one obtains $\sigma_{res}\simeq \sigma_{x}\simeq \sigma_{y}$. 

At an incoming flux similar to that used in the direct measurement of the spatial resolution ($200-300 \ \rm{Kparticles/s}$), we find $\sigma_{D}\simeq 7.7(4) \ t_{0}$ and thus $\sigma_{\text{res}}=2.7(2) \ t_{0}$. To compare this value to the direct measurement, obtained by imaging the grid, we evaluate $\sigma_{y}$ in the following way. The resolution $\sigma_{60}$ along the axis making a $\theta=60$ degrees angle with the $x$-axis reads $\sigma_{60}^2=(\cos(\theta) \sigma_{x})^2+(\sin(\theta) \sigma_{x})^2$. Using the measurement of $\sigma_{x}$ and $\sigma_{60}$ we obtain $\sigma_{y}=2.7(1) \ t_{0}$ and $\sqrt{(\sigma_{x}^2+\sigma_{y}^2)/2} =2.4(2) \ t_{0}$. This value is in good agreement with the spatial resolution $\sigma_{\text{res}}$ measured using the quantity $D$, demonstrating that one can indeed use a measurement of $\sigma_{D}$ to evaluate the spatial resolution from Eq.~\ref{Eq:ResFromSigmaD}. The usefulness of the quantity $\sigma_{D}$ lies in the fact that it provides a good estimate of the spatial resolution without the need to install any object under vacuum above the MCPs or to otherwise illuminate the detector with a specific pattern. 

The contribution of the electronic noise $\sigma_{{\rm t, noise}}$ to the resolution along the third-axis coordinate $t_{a}$ can be evaluated from the quantity $D$ in the same manner as for the in-plane resolution. From Eq.~\ref{Eq:tx1-tx2} we obtain that the RMS size of the distribution of $t_{a}$ is $\sigma_{t,a}=\sigma_{\text{res}}\simeq 0.3$ ns at 200 Kparticles/s. We note that this low value might not set the resolution along the third axis $t_{a}$ in certain experimental situations with incoming particles impacting the MCP at low velocities \cite{Bonneau2013}. 

Finally we investigate the influence of the incoming flux on the resolution $\sigma_{D}$. The values of $\sigma_{D}$ are plotted in Fig.~\ref{Fig:SigD} b) as a function of the incoming flux of particles. One sees that $\sigma_{D}$ varies significantly over the range of particle flux studied: while it is constant in the lowest 25\%, its value is multiplied by a factor 2 at 80\% of the maximum. A similar behavior has been observed with a lower threshold in a smaller ($40\rm{mm}$) chevron piled pair of MCPs combined with wedge and strip anode \cite{Jagutzki2002a}. This is related to the piling-up of pulses in the electronics and to the local gain depression of the MCP stack at high incoming flux \cite{Edgar1992}.

\section{Conclusion}

In this paper, we have presented the characterization of a complete detector chain based on a pair of MCPs and cross delay-lines in conjunction with a FPGA-based TDC we have recently developed. This detector provides access to the three-dimensional coordinates of single particles falling onto its surface at a high incoming rate. We have demonstrated that the architecture of our novel TDC permits a continuous acquisition of particles at  3.2 million particles per second, a performance that is significantly better than previously reported to our knowledge (for similar space and time detector). A further improvement of this rate will be provided by future TDC electronic developments which aim at doubling the multiplexer rate.  

We have measured the resolution of the detector by a traditional method (point spread function) and by using a quantity $D$ which yields similar results without the need to place any calibrated structure under vacuum. The simple use of  $D$ has allowed us to investigate the influence of the detected flux of particles on the resolution. In addition, this simpler alternative can be important when it is impractical to introduce a grid as for instance when the detector is under ultrahigh vacuum \cite{Bouton2015}.

\section{Acknowledgements}

We thank D. Boiron, H. Cayla, R. Lopes, P. Roncin and A. Villing for useful discussions and technical assistance. We acknowledge financial support from the R\'egion Ile-de-France (Grant Nano-K NewSPADE), the LabEx PALM (Grant number ANR-10-LABX-0039), the RTRA Triangle de la Physique and the Direction G\'en\'erale de l'Armement.



\begin{thebibliography}{21}
\bibitem{NoteSigmaX} Here $\sigma_{x}$ is defined in units of the coding time step $t_{0}$ of the TDC. Conversion to real spatial units is directly obtained from knowing the propagation speed of electronic pulses in the delay line \cite{RoentDek_Manual}.
\end{thebibliography}
\end{document}